\documentclass[sigconf]{acmart}
\usepackage{makecell}
\usepackage{multirow}

\renewcommand{\textcolor}[2]{#2}
\AtBeginDocument{%
  }

\copyrightyear{2026}
\acmYear{2026}
\setcopyright{cc}
\setcctype{by-nc-nd}
\acmConference[CUI '26]{ACM Conversational User Interfaces 2026}{July 21--24, 2026}{Bremen, Germany}
\acmBooktitle{ACM Conversational User Interfaces 2026 (CUI '26), July 21--24, 2026, Bremen, Germany}
\acmDOI{10.1145/3816046.3816301}
\acmISBN{979-8-4007-2741-2/2026/07}

\begin{document}
\title{After the Interface: Relocating Human Agency in the Age of Conversational AI}

\author{Mengke Wu}
\email{mengkew2@illinois.edu}
\affiliation{
    \institution{University of Illinois Urbana-Champaign}
    \department{School of Information Sciences}
    \city{Champaign}
    \state{Illinois}
    \country{USA}
}

\author{Mike Yao}
\email{mzyao@illinois.edu}
\affiliation{
    \institution{University of Illinois Urbana-Champaign}
    \department{Institute of Communications Research}
    \city{Champaign}
    \state{Illinois}
    \country{USA}
}

\renewcommand{\shortauthors}{Wu and Yao}

\begin{abstract}
As AI systems take on greater autonomy, a quiet anxiety has settled over the HCI community: human agency is eroding. Users no longer control execution, interfaces recede, and machines decide. We argue that this anxiety, while understandable, reflects a framing problem rather than an empirical finding. Agency has not diminished but has relocated. As interaction has shifted from command- and feature-based paradigms toward conversational, generative, and agentic AI, human agency migrates from interface affordances to interaction itself: articulating goals, evaluating outputs, and negotiating outcomes. To make this relocation visible, we revisit control as a diagnostic lens, distinguish process control and outcome control, and map different systems across this space to show that what looks like agency's disappearance is actually its redistribution. We take seriously the objection that outcome-based agency may be illusory in systems that produce plausible but unverifiable outputs, and argue that this concern reveals what agency in human-AI interaction truly requires. This paper invites the CUI community to reconsider what agency means, where it lives, and what it demands, including who gets to have it and who holds responsibility when it fails, before the consequences become impossible to overlook.
\end{abstract}




\maketitle

\section{Introduction}
The machine is taking over, or so the story goes. As conversational, generative, and agentic AI systems execute complex tasks with limited human oversight \cite{sun2024generative, SAPKOTA2026103599, naik2025designing}, a familiar worry surfaces in HCI research: users are losing control. When systems execute processes behind opaque interfaces and make decisions without explicit user input, traditional metrics suggest that human agency is eroding \cite{sundar2020rise, brandtzaeg2023good, heer2019agency}. This concern follows logically from how human agency has been made visible and measurable in HCI. For decades, agency was operationalized primarily through users' ability to control systems via visible interface affordances, \cite{norman1999affordance, holter2024deconstructing, inman2019beautiful}, evaluated by how many controls a system exposed, how transparent its internal states were, and whether users could intervene \cite{shneiderman2022human, wu2025negotiating, gebreegziabher2023patat}. This approach allowed designers and researchers to manifest and study agency through observable, static system attributes \cite{friedman1996value, friedman1996user}, and it embedded a consistent assumption: human agency scales with the degree of control a system explicitly affords. Within this framing, considerable research has focused on preserving human-in-the-loop, increasing transparency, and designing intervention mechanisms for autonomous systems \cite{wu2026rethinking, moruzzi2024user, lai2023towards, nakao2022toward}, which are valuable efforts that nonetheless implicitly accept a tradeoff in which preserving human agency requires constraining machine autonomy.

Yet this interpretation sits uneasily with how people engage with contemporary AI systems. Users may no longer control intermediate executions but still articulate goals, evaluate system responses, negotiate revisions, decide whether to accept or reject results, and determine when to abandon collaborations entirely \cite{gomez2025human, dhillon2024shaping, shi2023hci}. None of these actions map cleanly onto traditional interface affordance, but they are the actions through which people shape what AI systems produce and what they experience, or fail to experience, a sense of authorship and control. A growing mismatch exists between how agency has historically been operationalized and how it is actually enacted in contemporary human-AI interaction.

\textcolor{blue}{The central provocation of this paper is this: \textbf{human agency in AI systems has not eroded but has relocated, as what was once embedded in system affordances is now enacted through communicative interaction and outcome evaluation.} The apparent loss of agency is therefore not an empirical finding but a framing problem: the degree of agency users experience is no longer capped by what interfaces explicitly afford, but by how users choose to engage.} This shift has consequences for how we measure agency, design for it, and who gets to have it at all, which the field has not yet reckoned with.

To make this relocation visible, we revisit agency through the lens of control, a central operationalization of agency in system design and evaluation \cite{moruzzi2024user, bennett2023does, bergstrom2022sense, shneiderman2010designing}. Instead of treating control as unitary, we distinguish between \textbf{process control} (influence over methods and procedures) and \textbf{outcome control} (authority over the quality and acceptability of results). These two dimensions can vary independently, and in contemporary AI systems, they increasingly do. What looks like a loss of agency through the process lens may, through the outcome lens, represent a relocation into a space the field currently has not yet learned to see.

This relocation has consequences that reach beyond conceptual reframing. If agency now lives in communication rather than in affordances, our tools for measuring it are looking in the wrong place, our design frameworks are optimizing for the wrong thing, and the systems HCI community builds may be redistributing agency in ways we have not yet chosen to see. This paper traces that relocation, confronts its hardest objection, and asks what it would demand of how we study, design for, and take responsibility for agency in an era of increasingly autonomous AI.

\section{Background: The Evolution of Interaction Paradigms in HCI}
The relationship between human agency and system design has evolved substantially over the history of computing, reflecting how control, as an agency proxy, has been operationalized differently across interaction paradigms.

\textit{\textbf{Command-Based Interaction.}} In early computing, users exerted control through explicit and precise instructions, such as programming languages and command-line interfaces that required specifying every procedural step \cite{myers1998brief, shneiderman2010designing}. Users determined not only what outcome they wanted, but exactly how systems should produce it. Outputs were deterministic functions of inputs, and quality depended entirely on the precision of the instructions \cite{shneiderman1983direct, card2018psychology}.

\textit{\textbf{Feature-Based Interaction.}} The advent of graphical user interfaces shifted control toward feature manipulation \cite{shneiderman1983direct}. Rather than specifying algorithms, users clicked, dragged, selected, and adjusted through menus, buttons, sliders, or configuration panels. Procedural details automated, but sequential decisions about which features to activate remain with the user \cite{card2018psychology, cockburn2007predictive, myers2000past}. Systems like Photoshop or Excel afford considerable agency through the breadth of available features and their combinations. Outputs here remained relatively deterministic and traceable to explicit user actions \cite{Hutchins01121985}. Central to this paradigm was the assumption that affordances are visible: a button looks like a button, a slider communicates its range, and users can perceive what actions are available before taking them \cite{norman1999affordance}. Agency scaled with the legibility of the interface.

Over time, algorithmic mediation began to complicate this deterministic relationship. Recommender systems, search engines, and social media platforms allowed users to express preferences and provide feedback, but outputs were influenced rather than fully determined by user input, where the same settings could yield different results across time or users \cite{knijnenburg2012explaining, eslami2015always, lukoff2021design}.

\textbf{\textit{Conversational and Generative Interaction.}} Generative AI systems introduced a fundamental shift in how control is exercised. The affordance-based model encounters a rupture here: where a button communicates its function by its appearance, a conversational interface presents only an open input field. Users cannot perceive the system's capability boundary in advance, but only discover it through interaction \cite{kirschthaler2020can}. This is not a refinement of the feature-based paradigm but a structural break from it. Users now specify desired outputs through natural language without dictating procedural steps \cite{sanchez2024automating, gao2024taxonomy}, such as \textit{"Create a travel itinerary for Barcelona"} or \textit{"Summarize this document,"} while systems autonomously determine how to produce them. Users then can evaluate generated outputs and iteratively refine them through dialogue \cite{liu2023pre, zhou2024understanding} (e.g., \textit{"Add more details," "Try a different approach," "That's not what I meant"}). Beyond turn-taking interaction, increasingly agentic systems push this further: users specify high-level goals (e.g., \textit{"Increase social media engagement by 30\%," "Manage my mailbox"}) while systems plan and execute sequences of actions autonomously \cite{acharya2025agentic, SAPKOTA2026103599}, such as decomposing goals, coordinating tools, making interim decisions, and adapting strategies \cite{hughes2025ai, song2025multiagent}.

Users find themselves further from execution than ever, evaluating outcomes they did not witness being produced. This is the interaction paradigm that conversational user interfaces inhabit and where the tension between traditional agency frameworks and contemporary interaction is sharpest. Whether the judgment users exercise over outputs constitutes meaningful agency, and what it would take to find out, is what this paper sets out to examine.

\section{The Agency Paradox in Contemporary Human-AI Interaction} \label{paradox}

As these interaction paradigms evolved, so did approaches to understanding agency in HCI, yet the underlying assumption remained consistent. Agency is commonly understood as users' capacity to intentionally initiate action and experience oneself as influencing system behavior and outcomes toward desired goals \cite{zhang2025exploring, bennett2023does}. Across command-based and feature-based systems, users were understood to have greater agency when they could understand and proactively intervene in system behavior via visible interface affordances (e.g., specify execution steps, customize experiences, override defaults, provide feedback) \cite{wu2025negotiating, heer2019agency, sundar2020rise, gebreegziabher2023patat}. Agency was thus inferred from the presence, granularity, and accessibility of interface-level mechanisms that enabled these perceptions and actions \cite{moruzzi2024user, gebreegziabher2023patat}. This is a pragmatic operationalization shaped by the technical and interactional constraints of the time, where agency was naturally expressed through procedural control because that was how control could meaningfully be exercised in those systems \cite{holter2024deconstructing, shneiderman2022human}. It also allowed designers to reason about user empowerment through observable interface characteristics \cite{friedman1996value, friedman1996user}. The assumption embedded throughout was explicit: \textbf{human agency is bounded by the system and therefore scales with the degree of affordances a system exposes.} Users can only exercise as much agency as the system explicitly affords and at locations anticipated by designers. Within this view, as systems take on more decision-making or execution autonomy (e.g., automate steps, obscure internal processes, remove intervention points), they are often described as shifting control from users to machines \cite{sundar2020rise, brandtzaeg2023good, cavalcante2023meaningful}.

Contemporary AI systems alter this model fundamentally: users no longer control procedural steps through interfaces but instead specify desired outcomes and, in some cases, delegate entire workflows \cite{acharya2025agentic, SAPKOTA2026103599}. Viewing through the traditional logic, this creates a genuine paradox: as AI systems gain autonomy over processes, human agency appears to diminish, as users no longer exercise the granular control that historically defined their agency \cite{moruzzi2024user}. If agency is bounded by what systems afford, and systems are affording less procedural visibility, then agency is shrinking. \textbf{This conclusion, we argue, follows logically but from flawed premises.}

Norman's two gulfs offer a diagnostic lens here \cite{Hutchins01121985}. In prior paradigms, the Gulf of Execution was the primary site of friction, where users struggled to translate intentions into correct actions. Conversational and agentic AI largely dissolves this gulf: users speak or write, and systems act. Yet the Gulf of Evaluation deepens dramatically: users cannot observe what the system did, whether it succeeded in the sense they intended, or where it silently diverged. The hard problem has not been eliminated but migrated, and the field's current tools, calibrated for the old gulf, cannot see it.

The mismatch is not merely a measurement gap but reflects a shift in how agency is enacted. Even frameworks designed to address this tension, such as Shneiderman's Human-Centered AI that advocates for automation and human control simultaneously \cite{shneiderman2022human}, treat control as a single dimension, leaving unexamined what kind of control persists as systems grow more autonomous and where in the interaction it operates. In conversational and generative AI systems, users shape outcomes through evaluation, revision, negotiation, and selective acceptance. These forms of influence fall outside traditional notions of interface affordances but remain central to users’ sense of involvement, responsibility, and authorship \cite{fragiadakis2024evaluating, yu2025whose}. More fundamentally, agency in these systems is no longer determined solely at design time but is continuously emerging during interaction when users assess whether systems follow their intent, respect constraints, and behave in ways that feel appropriate \cite{shi2023hci, vereschak2024trust}. The problem, then, is not that agency has been reduced, but that it is now being enacted in the exchange itself, in a space the field's current tools cannot see.

But here the hardest objection arises. If agency has relocated to output evaluation, this assumes users can actually evaluate outcomes meaningfully. In probabilistic systems that hallucinate plausible but incorrect outputs, generate legally coherent but subtly flawed contracts, or produce confident-sounding but factually wrong responses, users may lack the capacity to detect failure \cite{tankelevitch2024metacognitive, buccinca2021trust, lee2025impact}. Is outcome-based agency, then, illusory?

We argue it is not, but the response requires reframing what evaluation means. The inability to fully verify outcomes is not a flaw unique to AI systems but is structurally present in human collaboration. When working with a collaborator or an assistant, we do not have process-level visibility into their cognition or methods and cannot independently verify every claim without prohibitive effort. Yet we do not conclude that agency disappears. Instead, agency becomes negotiated, relational, and trust-mediated, enacted through iterative exchange, accumulated credibility, and the ongoing exercise of judgment under uncertainty. \textbf{The shift from procedural control to output evaluation does not assume a perfect evaluator. It assumes that agency is enacted through negotiated judgment under epistemic uncertainty.} As AI systems become more autonomous, they increasingly resemble collaborators rather than deterministic tools, and the standard for agency shifts accordingly: from mechanical verification to relational trust, iterative negotiation, and the exercise of critical judgment over time.

This reframe has an uncomfortable corollary. \textbf{If agency is enacted through communicative negotiation rather than interface manipulation, it is no longer equally distributed by design.} Two users encountering the same conversational system can experience fundamentally different levels of agency, not because they access different features, \textcolor{blue}{but because they bring different communicative capacities, epistemic confidence, and persistence to the interaction, shaped by factors like personality, language background, domain expertise, and AI familiarity.} The articulate, persistent, and epistemically confident individuals who know how to prompt, push back, request revisions, and negotiate outputs gain more agency in this new regime. Those who do not, and passively accept the first response, lose it quietly without any interface feature changing. Unlike feature-based systems, where interface mastery translates into greater control, conversational systems offer no such ladder. The mapping between what a user says and what they get remains opaque and does not reliably improve with system familiarity. This is not a design failure in the conventional sense, as it cannot be fixed by adding an affordance or an override button. It is a shift in where human authority lives, and it raises questions about access, equity, and responsibility that the field has only begun to confront.

\section{Reframing Agency Through the Lens of Control: Process and Outcome}

To make the relocation of agency visible, we need a finer-grained view of control than the field currently uses. The gulf inversion described in Section \ref{paradox} (execution becomes frictionless while evaluation deepens) points toward what that view requires. Control in HCI research is often treated as a single dimension, where users have more or less of it \cite{wu2025negotiating, pu2011user, storms2022transparency, sieger2021exploring}. This made sense when control was exercised primarily through interface affordances, but it obscures important distinctions in contemporary AI systems: the difference between controlling \textit{how} something is produced and controlling \textit{whether the result is acceptable}.

We distinguish between two analytically independent dimensions. \textbf{Process control} refers to influence over methods, procedures, or steps through which outputs are produced, such as specifying workflows, constraining criteria, configuring parameters, or intervening during execution. \textbf{Outcome control} refers to authority over the quality and acceptability of results, such as assessing outputs against intentions and standards, iteratively refining through feedback and negotiation, or exerting judgment to accept, reject, or request modifications. These two dimensions can vary independently: process control does not guarantee acceptable results, and outcome authority does not require procedural visibility.

A critical distinction follows: output is not the same as outcome. A system may successfully produce an output (e.g., complete a task, generate a document, execute a workflow) without satisfying human intentions, quality standards, or values. Outcome control centers on judgment and evaluation beyond mere task completion. It is also distinct from the clarity or modality of user expression (e.g., prompting) but concerns what users can do with the results \textit{after} they are produced, not how precisely users can specify what they want beforehand.

\begin{figure*}
    \centering
    \includegraphics[width=1\linewidth]{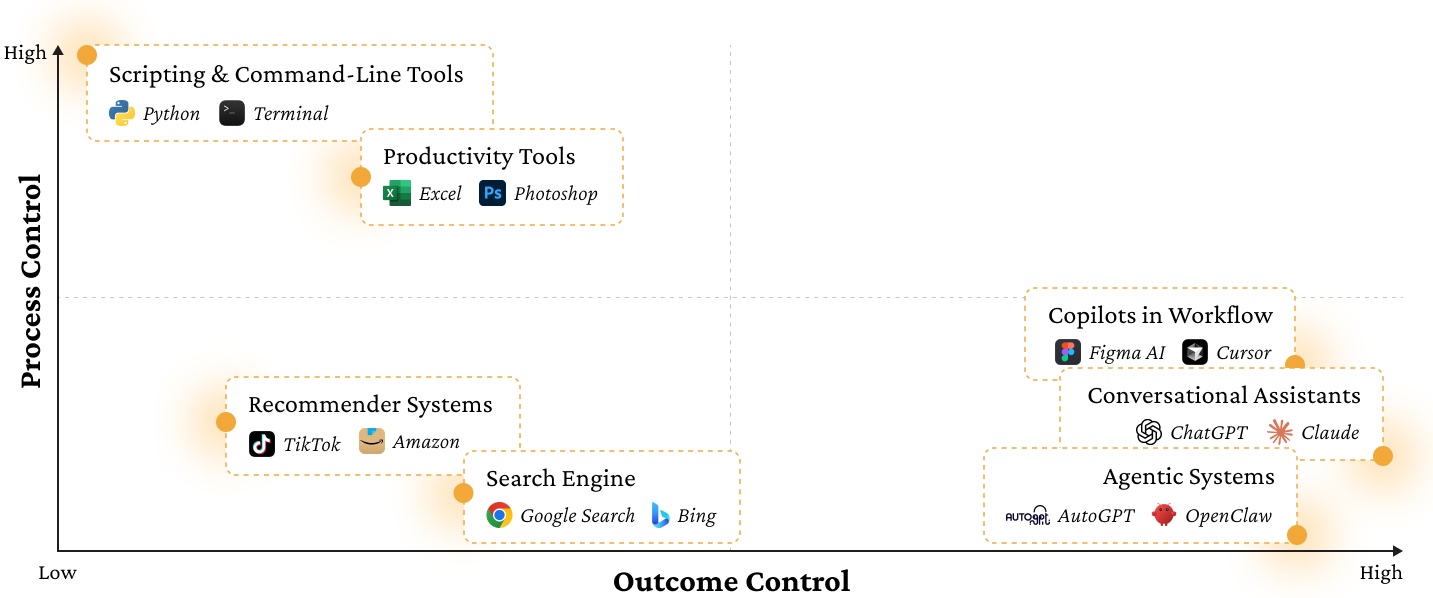}
    \Description{A two-dimensional scatter plot with Process Control on the vertical axis (Low to High) and Outcome Control on the horizontal axis (Low to High). Seven system categories are plotted: Scripting and Command-Line Tools (upper-left, very high process control, very low outcome control); Productivity Tools (upper-middle, mid-to-high process and low-to-mid outcome control); Recommender Systems (middle-left, low-to-mid process and low outcome control); Search Engines (lower-middle, low process and low-to-mid outcome control); Copilots in Workflow (right area, low-to-mid process and high outcome control); Conversational Assistants (far right, low process and very high outcome control); and Agentic Systems (lower-right, very low process and high outcome control). The three rightmost categories, Copilots, Conversational Assistants, and Agentic Systems, are enclosed in a dashed rectangle highlighting the high-outcome-control region. Each category is marked with recognizable product logos as examples.}
    \caption{A two-dimensional control space situating contemporary systems, with examples, by \textbf{Process Control} (vertical) and \textbf{Outcome Control} (horizontal). Positions are illustrative and not intended as complete or definitive classifications. Rationale for each system's placement is provided in Appendix \ref{rationale}.}
    \label{fig:map}
\end{figure*}

To make these dimensions concrete, Figure \ref{fig:map} maps contemporary systems into the two-dimensional space they define (see detailed list with rationale in Appendix \ref{rationale}). This mapping is not a taxonomy as the positions are illustrative, not definitive, and real systems often exhibit hybrid or shifting characteristics. Its purpose is diagnostic: to make visible how different interaction paradigms distribute control, and, in turn, shape experiences of agency. For example, scripting and command-line tools sit high on process control with limited interpretive flexibility: users specify exactly how, while the system decides nothing. In the middle lie familiar everyday systems, such as recommender systems that offer limited process control but also limited outcome negotiation: users influence but neither direct nor evaluate in any robust sense. Conversational and agentic systems sit at the far right, providing high outcome control with relatively limited procedural steering. This is the space the CUI community builds for, and it is precisely the space for which the field currently has the fewest tools (e.g., metrics, design patterns, frameworks) for understanding, measuring, and supporting the agency that lives there.

\section{What Does This Reframe Demand?}

\textbf{\textit{Measuring Agency.}} Existing measurements of perceived control, such as single-score surveys, feature checklists, and intervention point counts, capture process control reasonably well because they were designed to, but they have limited purchase on outcome control: on whether users are effectively shaping results, exercising meaningful judgment, or merely accepting outputs they lack the capacity to evaluate. Measuring agency in a conversational system would require attending to interaction trajectories rather than interface snapshots: when and why users push back, what triggers engagement shifts, how negotiation unfolds, and whether communicative moves translate into outcomes that reflect intentions. These are different conceptions of what agency measurement is for, not a refinement of existing methods. They require the field to decide what it is actually trying to measure, and for whom.

\textbf{\textit{Designing for Agency.}} Supporting outcome control is not simply a matter of enabling natural language feedback that most conversational systems already permit. The deeper challenge is scaffolding meaningful negotiation: helping users recognize that outputs are provisional, that revision is expected, and that their judgment matters. This might mean systems that proactively signal the negotiability of outputs (e.g., \textit{"Does this meet your needs?" "What should I revise?"}), provide structured affordances for complex feedback (e.g., section-level commenting, revision tracking, "keep this part, change that part" mechanisms), or make visible the gap between what was produced and what was intended. This also means confronting a more uncomfortable design question: if outcome control requires communicative competence, then designing for outcome control means designing for the unequal distribution of that competence. A design framework that ignores this inequality does not democratize agency but redistributes it toward those already advantaged.

\textbf{\textit{Accounting for Agency.}} Perhaps the most uncomfortable implication concerns accountability. As agency relocates from interface affordances to communicative evaluation, users bear increasing responsibility for outcomes they had decreasing visibility into producing. In high-stakes domains like legal, medical, and financial, accepting an AI-generated output without adequate evaluation is no longer simply a usability failure. It may constitute a failure of agency itself, with real consequences. Consider a user who instructs an agentic system to \textit{"clean up my inbox,"} the system completes the task with no errors, but deletes messages the user intended to keep. By process metric, the system succeeded; by outcome standard, it failed. Who bears responsibility lands squarely on the user: should they have been more specific? Should they have verified before the system acted? This is not an edge case but a structural pattern: the gap between literal instruction and intended outcome becomes a new site of failure, invisible to process-level metrics and unaddressed by design. It is also a site where evaluative agency is demanded of users by default. Yet the same systems that demand this evaluative agency from users are often the least transparent about what evaluation would require: what was generated, how, on what basis, and where the gaps lie. We are, in other words, redistributing responsibility upward to users at precisely the moment we are reducing their ability to discharge it. This paper does not resolve this tension. It insists the field cannot continue to ignore it.

\section{Conclusion}
Conversational AI systems have not diminished human agency, they have relocated it. As interaction shifts from procedural manipulation to communicative negotiation, human influence moves from the interface to the exchange itself. This paper has argued that this relocation is not merely a theoretical refinement but a practical necessity: our current tools for measuring agency, design frameworks, and assumptions about who gets to have agency are calibrated for a world that is rapidly receding. We invite the CUI community to take seriously what it means to build for agency when the interface is no longer where agency lives. The questions this raises about measurement, design, equity, and accountability do not have easy answers, but they are questions this community is uniquely positioned to confront.


\bibliographystyle{ACM-Reference-Format}
\bibliography{sample-base}

\clearpage
\appendix
\section{Appendix: Rationale for System Placement in the Two-Dimensional Control Space}
\label{rationale}

\vspace{4pt}
\noindent
\renewcommand{\arraystretch}{1.5}

\begin{tabular}{>{\raggedright\arraybackslash}p{3.1cm} >{\centering\arraybackslash}p{1.6cm} >{\centering\arraybackslash}p{1.5cm} >{\raggedright\arraybackslash}p{7.5cm} >{\raggedright\arraybackslash}p{1.9cm}}
\toprule
\textbf{} & \textbf{Process Control} & \textbf{Outcome Control} & \textbf{Rationale} & \textbf{Example} \\
\midrule
\textbf{Scripting \& Command-Line Tools} & Very High & Very Low & 
Users specify every procedural step with complete precision, but outputs are deterministic functions of inputs with no room for reinterpretation, negotiation, or iterative quality refinement & 
Python, Terminal \\
\midrule
\textbf{Productivity Tools} & Mid to High & Low to Mid & 
Users control execution through extensive features that precisely define workflows. There is some space for iteration in the results, but they are mainly deterministic and not subject to interpretation or negotiation & 
Excel, Photoshop \\
\midrule
\textbf{Recommender Systems} & Low to Mid & Low & 
Users have no direct control over ranking algorithms or recommendation logic, and provide only indirect influence through engagement and preference signals. Their ability to shape outputs is limited to passively accepting or rejecting items, bypassing via search, or comparing alternatives without negotiation & 
TikTok, Amazon \\
\midrule
\textbf{Search-based Systems} & Low & Low to Mid & 
Users cannot control ranking algorithms or retrieval logic, but can explicitly specify goals through queries and evaluate returned results, then iteratively refine searches to improve results & 
Google Search, Bing \\
\midrule
\textbf{Copilots in Workflow} & Low to Mid & High & 
Users partially steer execution by selecting when, where, and how the copilot intervenes, then review, accept, reject, or refine suggestions with full awareness of the surrounding work context & 
Figma AI, Cursor \\
\midrule
\textbf{Conversational Assistants} & Low & Very High & 
Users cannot control internal reasoning or execution, but can continuously evaluate outputs, negotiate revisions across unlimited turns, reframe goals mid-interaction, and exercise iterative judgment with no structural endpoint & 
ChatGPT, Claude \\
\midrule
\textbf{Agentic Systems} & Very Low & High & 
\textcolor{blue}{Users set high-level goals while systems autonomously plan and execute multi-step workflows; users then evaluate whether goals are achieved and can redirect or abandon. However, unlike conversational assistants where every output is immediately visible and negotiable turn-by-turn, agentic systems' execution opacity and absent continuous negotiation partially constrain users' ability to detect divergence and redirect in real time} & 
AutoGPT, OpenClaw \\
\bottomrule
\end{tabular}
\vspace{4pt}

\end{document}